# Application of Markov Chains to Multiple Sclerosis Clinical Trial Data to Estimate Disease Trajectories

Uma Sthanu, Gary Cutter

## Abstract

**Background**

Multiple Sclerosis (MS), an autoimmune disease affecting millions worldwide, is characterized by its variable course, in which some patients will experience a more benign disease course and others a more active one, with the latter leading to permanent neural damage and disability.

**Methods**

This study uses a Markov Chain model to demonstrate the probability of movement across different states on the Expanded Disability Status Scale (EDSS) and attempted to define worsening, improvement, cycling, and stability of these different pathways. Most importantly we were interested in assessing the lack of impermanence of confirmed disability worsening and if it could be estimated from the Markov model.

**Results**

The study identified only 8.1% were considered worsening, 5.6% consistent improving and 86% cyclers and less than 1% consistently stable. More importantly we also found that many (approximately 30%) of participants with confirmed disability worsening (CDW) regressed to stages that were not considered worsening, on subsequent visits after CDW.

**Conclusions**

These finding are similar to what has been reported previously as predictors of worsening, and also for a lack of durability of CDW, but our results suggest that clinical trial endpoints may need to be modified to more accurately capture differences between the treatment and control groups. Further, this suggests that the rate of worsening in trials that use time to CDW are overestimating the extent of CDW. The trials remain valid since the regressing applies to both treatment and control groups, but that the results may be underestimating the treatment benefit due to misclassification.

# Background

Multiples Sclerosis results in a wide range of phenotypes across patients. While personalizing MS disease treatment may lead to more positive outcomes for patients, such tailored treatment is difficult without specific prognostic factors to predict what disease course individual participants will have, and whether they are likely to experience a progressive course, or more benign course.

Magnetic Resonance Image (MRI) Lesions have, in the past, been considered a possible surrogate endpoint in clinical trials to denote disease activity in Multiple Sclerosis (MS) participants - however, their use on an individual level has not been widely correlated to disease outcomes, and this study aimed to assess their value as well as others in early MS participants as predictive measures of future disability [1].

Functional tests such as the 9-hole Peg Test and the Timed 25-Foot Walk have, similarly, been used as endpoints and have been found to possess monitoring capabilities as part of the Multiple Sclerosis Functional Composite (MSFC) score [2,3,4,5]. This study aimed to use baseline readings from these two tests for participants to determine whether a patient's early readings may suggest their future disease course as opposed to predicting results at landmark time points alone.

# Methods

The 1,008-patient CombiRx Clinical Trial dataset was used for this study [6]. To characterize the trajectories and patterns of participant experience, a Markov chain was employed. A Markov Chain is a probabilistic process in which individuals move from one state to another or stay put in a state, the probabilities of which only depend on the current state. This so-called memoryless property allows predictions which in essence ignores how one got to a state. This approach was supported by Confavreaux [7] in describing the common trajectory of MS patients after reaching a disability status of 4 independent of their history of relapses. Confavreaux argued that the history of relapses did not essentially influence the irreversible disability in the long term. Thus, maybe the pattern of movement from one EDSS state to the next becomes the determining factor. The concept of progression independent of relapses (PIRA) certainly has its origins in the concepts proposed by Confavreaux, despite the many problems of implementing the PIRA concept.

Two hundred randomly selected participants were set aside for validation for the Markov chain portion of the study, with 808 set aside to build the time course of MS by specifying movement from EDSS state to EDSS state at three-month intervals, the transition matrices from the beginning of the study. While the dropout rate in CombiRx by current standards was very low (19% at three years), nevertheless, due to dropouts, the number in each of these two cohorts (discovery and validation) steadily decreased by the end of the study, with the validation cohort ultimately possessing 159 participants (79.5%) after 36 months.

In the first phase of the study, the Markov chain transition matrices were created for each time point (month 0 to month 3 was one matrix, month 3 to month 6 was another, etc.). Each matrix, in accordance with the setup of a Markov chain, was filled with the probability of a participant transitioning from being at the state identified as the row variable at a specified monthly time point to the column state representing where the participant was at the ending time point 3 months later, given that the patient was in the row state at the start of the interval, known as the transition probability matrix and mathematically we write:

$$M_{ij} = P(j \mid i)$$

Where $M_{ij}$ represents the individual elements of the transition matrix, j is the column (future) state, and i is the current (immediately preceding j) state utilizing the trial structure where assessments were made every three months. Unscheduled visits were ignored.

Disability state groups were defined based on the EDSS groupings: 0, 1-1.5, 2, 2.5, 3, 3.5, 4-5.5 and >=6. This was done to account for the entry criteria of participants in the study not being permitted to have an EDSS state above 5.5, and the states of 6 or higher were treated similarly although this population was treatment naïve at the start of the study and unlikely to progress beyond 6.5 in the three years under consideration.

These transition matrices were treated in three different ways in order to determine their predictive ability in modeling the course of disease:

1. The baseline distribution of participants across the state groups in the discovery cohort was defined as the "baseline distribution" and set as the expected distribution (**Table 2**). It was then compared with the distribution of a validation cohort at every time point in the study.
2. Next, the transition matrices at each 3-month visit (created based off the discovery cohort) were averaged across all time points using the 12 transition matrices from month 0 to month 36. Also, a property of the Markov chain, called a memoryless property or stationarity, where an individual's future state depends only on their current state, allows for future predictions once probabilities have been computed for an individual transition matrix. This property allows the future distribution to be found by combining one transition to the next transition matrix by multiplying the two matrices and beyond (i.e. chaining), resulting in raising the estimated transition matrix to the power of a future number of transitions. When assumptions are met, the matrix reaches a so-called steady state and is used as the expected distribution at each time point. For example, to find the expected distribution at Month 6 (2 time intervals), the distribution anticipated by raising the transition matrix M to the second power was used, or M x M given two transitions had passed (Month 0 → Month 3, and Month 3 → Month 6). This created what is hence referred to as the homogeneous or averaged matrix.
3. Finally, the separately estimated transition matrices from the discovery cohort were multiplied together at each time point to create transition probabilities that changed at each time point. For example, to get the expected distribution at Month 6, the transition matrix where the rows were Month 0 states and the columns were Month 3 states was multiplied by the transition matrix that went from Month 3 to Month 6, hence deriving the probability of transitioning from each state group to each other from Month 0 to Month 6. This created probabilities that varied at each time point, hence referred to as the inhomogeneous matrix.

Each expected distribution, based on the methods described above from the discovery cohort, was multiplied by the number of participants (after dropouts) remaining in the validation cohort at each time point to come up with an expected count of participants in each state for the validation cohort. These numbers were then compared with the actual number of participants in the validation cohort in each state and assessed using a chi-squared goodness of fit test.

Following the creation of the Markov Chain, four paths were defined for different disease courses that a patient might experience on the chain (initial characteristics in **Tables (3-4)**):

1. Worsening: Participants who had at least one instance where they were above their EDSS baseline reading and did not have any cases in which they went below their EDSS baseline reading were considered Worsening.
2. Improvement: Participants who had at least one instance where they were below their EDSS baseline reading and did not have any cases in which they went above their EDSS baseline reading were considered Improvers.
3. Stability: Participants who remained in their baseline EDSS state throughout the course of the study were considered Stable.
4. Cycling: Participants who had instances of both improving from their baseline EDSS reading and worsening from their baseline reading were considered Cyclers.

CombiRx followed participants until the last participant finished - thus, 3 to 7 years. In order to minimize the impact of how long the participant was in the study on their probability of experiencing each disease course, the classifications of each participant into the four disease courses were restricted to being defined by their disease states over three years only, and participants who dropped out before three years are considered as outcomes in the Markov chain portion comparing the fit of the models. Chi-square goodness of fit statistics summarizing the comparison of the observed data to the expected were computed to assess the fit of the models and used as descriptive statistics.

Based on these results we estimated the progression rates by EDSS level at baseline and further assessed how many participants with confirmed disability worsening subsequently regressed to levels below the threshold for determining progression.

We generated model-predicted probabilities of Confirmed Disability Worsening (CDW) and the model based expected regression from CDW. Regression were found by finding the probability of possible "walks" on the averaged Markov chain that identified moving to levels at or below the initial threshold of worsening which defines progression (generally 1 step on the EDSS scale) We accomplished this calculation by iteratively updating disability state probabilities over multiple transitions. This was done repeatedly to allow us to find the probability of regressing by the last state the participant was in when they achieved CDW.

A participant was marked as "progressing" based on the definition of 6-month CDW, and "regressors" were those that went back below the EDSS state that triggered CDW at least once by the time of the last observation that participant, and this value was no more than one EDSS state above baseline thereby indicating regression to a pre CDW level of disability. For example, a participant who start at EDSS = 2 and experiences a progression at 9 months and meets the definition of CDW with EDSS scores of 4, 3, 3 for months 9, 12 and 15 respectively, but then at month 18:

If the EDSS were 3 this would not be regression since the disability level remains above the threshold for an initial EDSS of 2 even though it is below 4 the value initiating their CDW.
If month 18 had been an EDSS= 2.5 then this would be regression because a value of 2.5 is below the threshold of EDSS=3 and thus, would no longer qualify as progression and is considered here as regression.

# Results

The discovery and validation cohorts and their descriptive statistics are shown in **Table 1**. The two cohorts were comparable with respect to the covariates used in the evaluation of the fit of the modeling. They were of mean age 37.7 with an EDSS of slightly over 2, 72.4% female, 1.2 years since diagnosis, and a consistent and normal level of walking and 9HPT values with low lesion burden. Table 2 shows the distribution of the EDSS scores with a medium disability of 2 and 75% of the population below 3.

Chi-Squared goodness of fit values were tracked over time for each of the created Markov Models and their respective predicted distributions at each time point based on the discovery cohort, then compared with the actual distribution of the validation cohort (**Figure 1**). Using the baseline distribution to predict subsequent disability status showed increasing chi-square values as might be expected indicating participants were not essentially staying stable for 3 years (Figure 1A). This severely underestimated the number of participants who would end in the 6-10 EDSS states since the cohort had only 1 participant in this category, given the entry criteria of the clinical trial used. Figure 1B shows the results of the averaging of the 12 monthly matrices to predict each observation time distribution and this tended to fit the data more consistently except at 6 months. To assess if this underestimation leading to lack of fit was indeed due to early results, the analysis was re-performed only considering the matrices after one year. This inhomogeneous Markov chain was then found to converge on a more accurate stationary distribution (Supplementary **Table 1, Figure 1C**)

This approach led to a stationary distribution and was found to have a goodness of fit chi-squared value of 10.23 after 48 months with 7 degrees of freedom ($p>0.05$), meaning a non-significance between the observed and expected counts. While imperfect, this still suggested a reasonably close fit of the observed data to the predicted after 1 or more years of treatment. The assigned proportions are shown over time in Supplementary Figure 1.

When examining the classification or trajectories, 8.1% of participants were classified as worsening; 5.6% as improvers; 86.0% as cyclers and less than 1% were considered stable being defined with the same EDSS values throughout.

We next examined the predicted probabilities of CDW from the averaged Markov Chain and compared them to the actual number progressing. Figure 2 shows the proportion of individuals with CDW (green bars), the model percent predicted to experience regression to a lower EDSS level given they achieved CDW and the Actual proportions of those with CDW to that EDSS level who subsequently regressed below their CDW minimum threshold. Table 3 show the agreement between the predicted regression and the observed regression. Except for those who achieved CDW at the 1-1.5 EDSS level, the modeled probabilities of regression are not significantly different from the observed proportions. The regression from EDSS 1-1.5, showing no regression may be due in part to mismeasurement of the EDSS at baseline. For example, if a participant was erroneously classed at EDSS=0 at baseline, then early progression at 3 months as CDW would simply be do to correctly measuring them at 1-1.5 subsequently and thus, regression not occurring. The sample sizes shown in the table are relatively small indicating the low numbers of CDW over the 3 year period ranging from 14 to 50. The rates of regression approximating 30 percent differ slightly by EDSS level but are shown both by expected and observed values to be more at the higher ends of the EDSS scale, whereas we expect more measurement error at the lower end of the disability scale.

# Discussion

This study is not the first to use this methodology to assess the course of MS, but it is the first to look specifically consider specific disease courses, as we defined. First, the Markov chain methodology was found to be consistent with the assumption following the Markov memoryless property and to model the MS clinical trial EDSS scores for a validation cohort with reasonable accuracy, as exemplified by the non-statistically significant Chi-Squared values summarizing the goodness of fit of the model over time and the convergence of the chain to a stationary distribution especially after ignoring the initial early assessments which may have been disproportionately contributing to the model and may be evidenced by the lack of regression from the 1-1.5 EDSS category.

Confavreaux [7] was one of the first to point out that relapses were not associated with progression after a disability status of 4, but he did not consider the potential ups and down in the DSS (disability status scale, a 1 point interval scale that may have had less variability than the EDSS). Earlier clinical trials used the concept of 5 or more lesions, the Rule of 5 [9] to monitor for excess risk at the participant level in trials, but quite possibly lower levels of gadolinium activity are leading to changes in disability with consequences leading to the cycling of EDSS levels and may be further contributing to false positive disability classification and may help explain the durability and lack thereof found by Satyanarayan [10].

From the models that ignore the first 3 months on treatment suggest over the course of two and a half **years a majority of patients experienced cycling behavior in their EDSS states**. This could be indicative of the activeness of the disease course of these patients, given that they were newly diagnosed and naïve MS patients at the beginning of the study. However, it could also be indicative of measurement error and the inherent variability of EDSS, underlying biological variability, an issue that continues to demand the creation, validation, and acceptance of new measures of MS disability.

As MS has consistently moved to use time to event models for developing and comparing therapeutic treatments, these results have marked importance. Within a treatment group using time to event clearly overestimates the permanent disability that is occurring. Randomized clinical trials prevent this by having control groups that also experience this regression phenomena. However, the actual risk reductions may be greater than seen because the potential false positive CDW participants are contained within both treatment groups. On the flip side, it means that the proportion actually worsening is overestimated. It should be noted that there have been many improvements in the assessment of the EDSS in the time since the CombiRx trial was conducted. The Neurostatus has improved the consistency of the measurement and probably has removed some of the regression that is due to measurement error. Nevertheless, the modeling approach demonstrated here could be used to better estimate the actual disability accrual in a trial population.

# Limitations

This study is not without limitations. It sought to identify longer term trajectories of worsening and improvement. A potential limitation is the fact that CombiRx was a treatment trial with no placebo control group. While common now, we are looking at the trajectories of participants on treatment, and thus, the predictions are within the context of treatment and injectable treatments which could have influenced both progression and regression, although we have no data to support that skipping injections leads to EDSS progression and more importantly that resuming such injections would suddenly improve prior assessed disability, although theoretically possible.

CombiRx did not identify important differences amongst the three treatments and thus, treatments were ignored in these analyses. Nevertheless, these predictions of disease course were found in the face of these older first-generation treatments and the better fit of the models using the first three months data, may actually be a treatment effect of who may or may not respond to these first generation agents. The unexpected lack of the EDSS regression at the 1-1.5 level of progression seems more likely early measurement error or variability than individuals at lower levels of the EDSS less likely to respond to treatment. While not a limitation in and of itself, others have shown that when new lesions occur while someone is on treatment, this may be specific to the treatment and/or indicative of treatment failure rather than the natural history of the disease, although this could have caused some CDW as well as regression hidden by longer recovery when these relapses occur in a treated population.

# Conclusion

This study used a discovery cohort that was independent of the randomly selected validation cohort using MS clinical trial data. It demonstrated that the process can be modeled with a memoryless process, and a time-inhomogeneous Markov model was built in order to model the movement of Multiple Sclerosis participants across different state groups as defined by the Expanded Disability Status Scale. The Markov

Model built upon different transition matrices for each time point yielded higher accuracy than one averaged across all time points, although these gains were reduced by the end of three years. The above findings suggested that, as is expected in MS, even under treatment, the cohort as a whole shifted its distribution over time and showed that the cohort approached a stationary distribution that was distinct from its initial distribution but was relatively stable over the time frame considered.  This implies that some who transition out of various EDSS states are replaced by others transitioning into that state. This should cause some concern with the concept of confirmed disability worsening, although presumably unbiased when comparing two treatment groups, may overstate the benefit in that the label conveys the concept of permanent damage, which may indeed be transitory.

The high rates of regression or so called cycling behavior argues for more complex study using the Markov approach to help identify when the transition from this relapsing remitting phase truly becomes a progressive phase and what variables might be predictive of this transition.  Our findings suggest that progression patterns shown as short-term disease behavior, rather than long- or medium-term disease activity and understanding more about this regression behavior is important as many studies continue following a treated cohort in an open label extension study and the regression may be misinterpreted as longer term improvements from the drug being utilized, when it is from a concept of confirmed disability worsening which is taken as non-recoverable disability.

# Tables & Figures

**Table 1:** Baseline characteristics of the cohorts used during the Markov Chain process.

|  |  | Discovery (n=808) | Validation (n=200) | Total |
|---|---|---|---|---|
| Baseline EDSS State **Average (St. Dev, median, IQR)** | | 2.1 (1.2, 2, 1.125) | 2.3 (1.1, 2, 1.5) | 2.1 (1.2, 2, 1.5) |
| Age **Average (St. Dev)** | | 37.7 (9.61) | 37.9 (9.84) | 37.7 (9.7) |
| Sex **(n)** | F | 586 | 144 | 730 |
|  | M | 222 | 56 | 278 |
| Years since diagnosis **Average (n, St. Dev)** | | 1.2 (3.2) | 1.2 (3.8) | 1.2 (3.3) |
| 9HPT Average **Average (St. Dev)** | | 20.2 (5.0) | 20.5 (4.4) | 20.3 (4.9) |
| T25FW Average **Average (St. Dev)** | | 4.9 (1.6) | 5.5 (1.7) | 5.0 (1.6) |
| 1/9HPT Average **Average (St. Dev)** | | 0.1 (0.0) | 0.1 (0.0) | 0.1 (0.0) |
| Gad Lesion Count **Average (St. Dev, median, IQR)** | | 1.5 (3.7, 0, 1) | 2.7 (5.9, 0, 3) | 1.7 (4.2, 0, 2) |
| T2 Lesion Volume **Average (St. Dev, median, IQR)** | | 10.2 (11.0, 6.3, 10.6) | 11.7 (13.3, 6.8, 12.3) | 10.5 (11.5, 6.4, 10.8) |
| T2 Lesion Count **Average (St. Dev, median, IQR)** | | 86.7 (56.6, 70.0, 60.0) | 93.2 (53.4, 80.5, 70.8) | 88.0 (56.1, 71.0, 62.3) |

**Table 2:** Baseline distribution of cohort used as baseline model

| State group | 0.0 | 1.0-1.5 | 2.0 | 2.5 | 3.0 | 3.5 | 4-5.5 | 6-10 |
|---|---|---|---|---|---|---|---|---|
| Portion of population | 0.110 | 0.254 | 0.254 | 0.132 | 0.103 | 0.072 | 0.074 | 0.001 |

Table 3

| EDSS at CDW | Expected | Observed | N | | t-value | p-value |
|---|---|---|---|---|---|---|
| 1-1.5 | 0.173 | 0 | 43 | | -2.9991938 | 0.0045 |
| 2 | 0.232 | 0.2791 | 50 | | 0.78900782 | 0.4339 |
| 2.5 | 0.256 | 0.38 | 27 | | 1.47637648 | 0.1519 |
| 3 | 0.242 | 0.2593 | 28 | | 0.21373854 | 0.8324 |
| 3.5 | 0.197 | 0.3214 | 17 | | 1.28959683 | 0.2155 |
| 4-5.5 | 0.376 | 0.2941 | 23 | | -0.8108894 | 0.4261 |
| 4-4.5 | 0.333 | 0.3043 | 13 | | -0.2195677 | 0.8299 |

**Supplementary Table 1:** Stationary Distribution with underestimation versus stationary distribution with omitted year

| State group | 0.0 | 1.0-1.5 | 2.0 | 2.5 | 3.0 | 3.5 | 4-5.5 | 6-10 |
|---|---|---|---|---|---|---|---|---|
| % of population predicted with underestimation | 12.4% | 27.8% | 23.1% | 12.1% | 10.4% | 5.4% | 6.3% | 2.4% |
| % of population predicted with omitted year | 11.9% | 26.9% | 22.6% | 12.2% | 10.7% | 5.6% | 7.0% | 3.2% |

**Figure 1:** Chi-squared values of models described in Methods over time

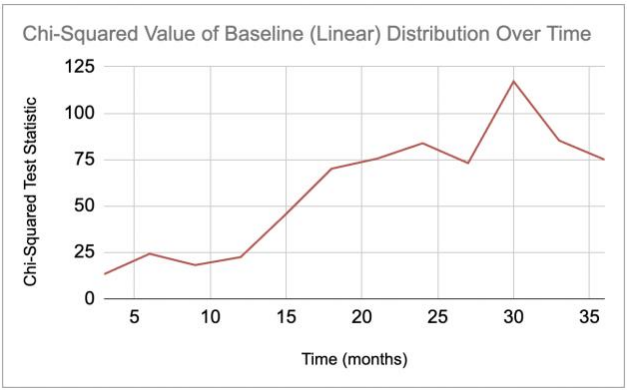

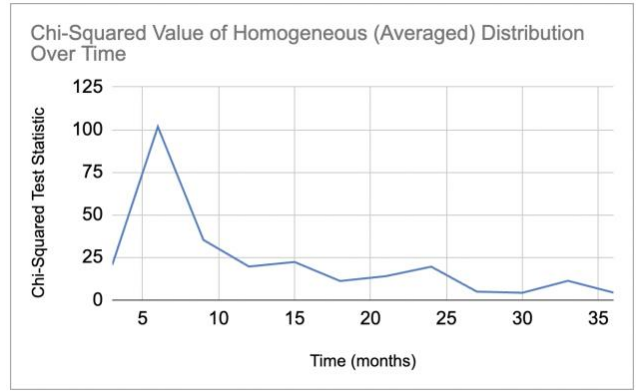

A. Baseline distribution used to predict distribution

B. Average of the 12 transition matrices to predict distribution

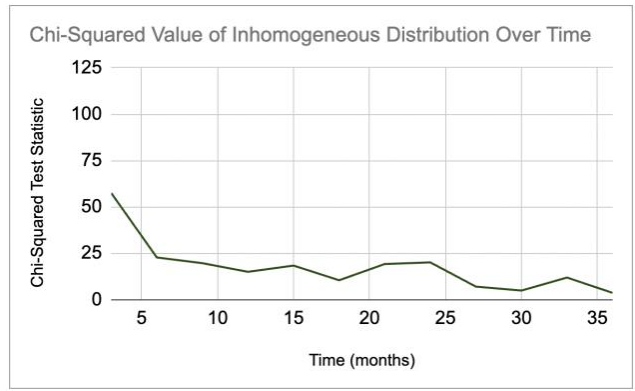

C. Individual transition matrices to predict distribution

**Supplementary figure 1:** Probability of being in each EDSS state group as predicted by inhomogeneous chain over time

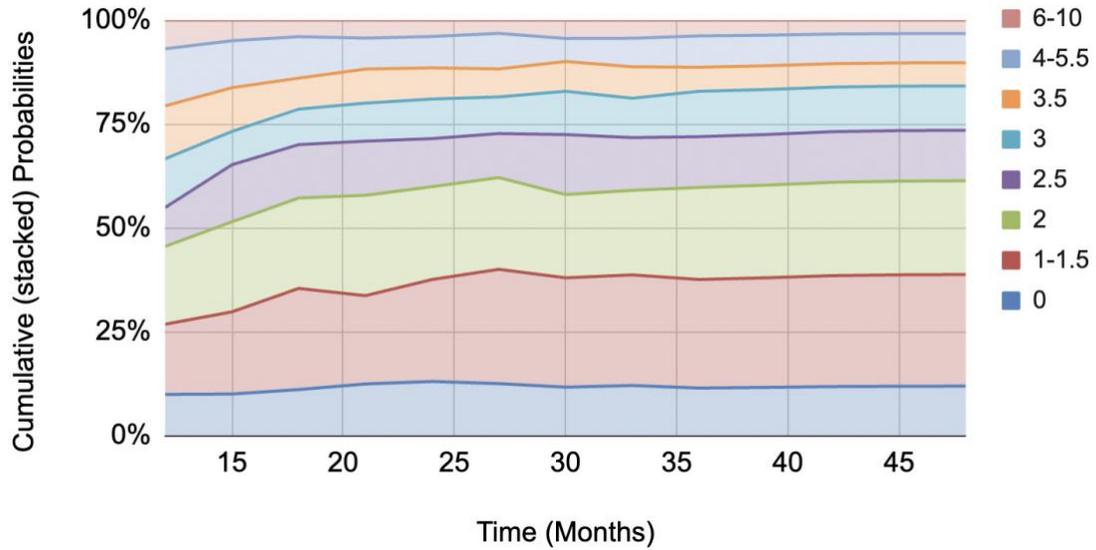

Figure 2 **Proportion Predicted to meet CDW and Proportion Predicted Regressing from CDW and Actual Proportion Regressing**

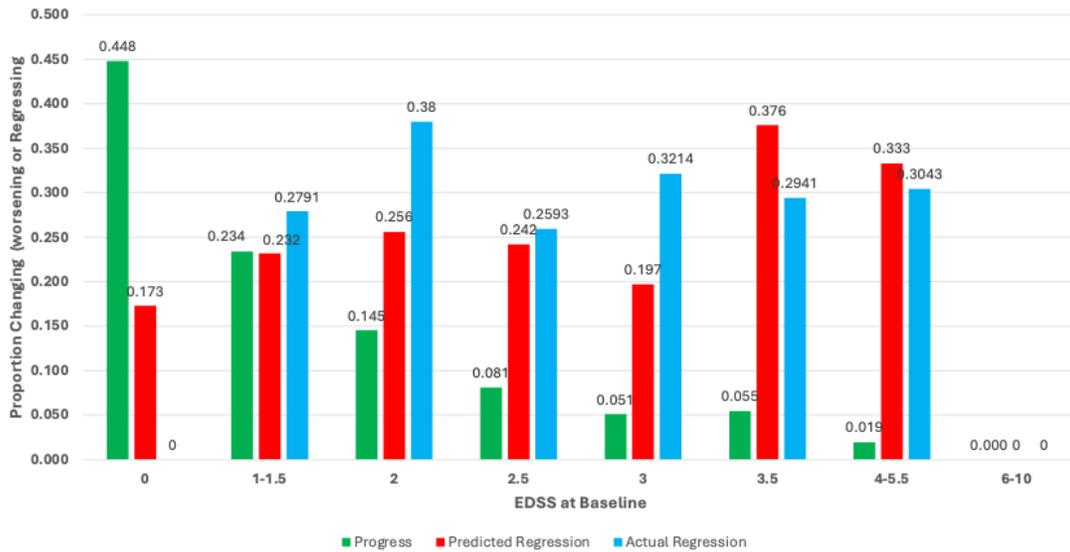